\documentclass[a4paper,11pt]{article}
\pdfoutput=1 

\usepackage{jinstpub} 

\title{\boldmath New prototype scintillator detector for the Tibet AS$\gamma$ Experiment}

\author[a]{Y. Zhang,}
\author[a]{Q.-B. Gou,}
\author[b,a]{H. Cai,}
\author[c]{T.-L. Chen,}
\author[c]{Danzengluobu,}
\author[d]{C.-F. Feng,}
\author[a,e]{Y.-L. Feng,}
\author[a]{Z.-Y. Feng,}
\author[c]{Q. Gao,}
\author[a]{X.-J. Gao,}
\author[a]{Y.-Q. Guo,}
\author[a,e]{Y.-Y. Guo,}
\author[a]{Y.-Y. Hou,}
\author[a,e]{H.-B. Hu,}
\author[a,e]{C. Jin,}
\author[c]{H.-J. Li,}
\author[a,1]{C. Liu,\note{Corresponding author.}}
\author[c]{M.-Y. Liu,}
\author[a,f]{X.-L. Qian,}
\author[a,e]{Z. Tian,}
\author[a,e]{Z. Wang,}
\author[d]{L. Xue,}
\author[d]{X.-Y. Zhang}
\author[a]{and Xi-Ying Zhang}

\affiliation[a]{Key Laboratory of Particle Astrophysics, Institute of High Energy Physics, Chinese Academy of Sciences, \\Beijing 100049, China}
\affiliation[b]{School of Nuclear Science and Technology, Lanzhou University, \\Lanzhou 730000, China}
\affiliation[c]{Physics Department of Science School, Tibet University, \\Lhasa 850000, China}
\affiliation[d]{Shandong University, \\Jinan 250100, China}
\affiliation[e]{University of Chinese Academy of Sciences, \\Beijing 100049, China}
\affiliation[f]{Shandong Management University, \\Jinan 250100, China}

\emailAdd{liuc@ihep.ac.cn}

\abstract{The hybrid Tibet AS array was successfully constructed in 2014. It has 4500 m$^{2}$ underground water Cherenkov pools used as the muon detector (MD) and 789 scintillator detectors covering 36900 m$^{2}$ as the surface array. At 100 TeV, cosmic-ray background events can be rejected by approximately 99.99\%, according to the full Monte Carlo (MC) simulation for $\gamma$-ray observations. In order to use the muon detector efficiently, we propose to extend the surface array area to 72900 m$^{2}$ by adding 120 scintillator detectors around the current array to increase the effective detection area. A new prototype scintillator detector is developed via optimizing the detector geometry and its optical surface, by selecting the reflective material and adopting dynode readout. This detector can meet our physics requirements with a positional non-uniformity of the output charge within 10\% (with reference to the center of the scintillator), time resolution FWHM of $\sim$2.2 ns, and dynamic range from 1 to 500 minimum ionization particles.}


\begin{document}
\maketitle
\flushbottom

\section{Introduction}
Since the discovery of cosmic rays (CRs) in 1912, their origin remains a fundamental problem. Many astrophysical objects, such as supernova remnants (SNRs) \cite{MNRAS1978_182_147,MNRAS1978_182_443,ApJ1978}, galactic center (GC) \cite{AZh,NJPh2013,GUO2017}, and others \cite{PRL1995,APP1999}, are supposed to generate expanding diffusive shocks and accelerate CRs to a very high energy. At current status, to observe $\gamma$-rays is the optimal way to identify which object makes the dominant contribution. In recent years, great progress in the measurement of the $\gamma$-rays have been made with new generation of space borne and ground-based experiments. The Fermi large area telescope (LAT) observed the characteristic pion-decay feature in the $\gamma$-ray spectra of three SNRs, IC 443, W44, and W51C \cite{Science2013,Apj2016}. This detection provides direct evidence that CR protons are accelerated in SNRs. Peta-electron volt CRs, likely being accelerated by the GC, recently have been reported by the HESS experiment \cite{Nature2016}. Though all these discoveries shed some light on the problem of the CR origin, the dominant sources of the contribution to galactic cosmic rays (GCRs) is still unclear, and it is necessary to discover new hadronic sources. Deu to fast energy loss of an electron by synchrotron radiation and inverse Compton scattering with a background photon and the Klein--Nishima effect, a hard $\gamma$-ray spectrum in the 100 TeV energy range is believed to originate from the decay of $\pi^0$ generated in the interaction of very high-energy CRs with the ambient gas \cite{PRD1995}. Therefore, the observation of $\gamma$-ray sources in the 100 TeV energy range is one of the best ways to resolve the problem of GCR origin.

With the advantages of both high altitude and wide field of view, the Tibet AS$\gamma$ experiment was started in 1990 at Yangbajing (90$^{0}$31$^{'}$ E, 30$^{0}$06$^{'}$ N; 4300 m above sea level) in Tibet, China. The Tibet AS experiment currently has the surface array with 789 scintillator detectors covering an area of 36900 m$^{2}$ and the 4500 m$^{2}$ underground water Cherenkov pools used to select muons\cite{lab2,TeV100,MDA}. The Monte Carlo (MC) simulation predicts that CR background events will be rejected by approximately 99.99\% at 100 TeV using this upgraded hybrid experiment \cite{lab3}. In order to use the muon detector efficiently and to increase the effective detection area, one upgrade is proposed by adding 120 scintillator detectors around the current array. The area of the surface array can be extended to 72900 m$^{2}$ according to this proposal.

Scintillator detectors are widely used in extensive air shower (EAS) experiments (e.g., Tibet AS$\gamma$ \cite{lab4}, KASCADE-Grande \cite{kascade}, GRAND \cite{grand}, and so on). For the Tibet AS$\gamma$ array, scintillator detectors have been running stably for more than 25 years. In the large high-altitude air shower observatory (LHAASO), it is also planned to build 5195 scintillator detectors, each with an area of 1 m$^{2}$, as electromagnetic particle detector (ED) array \cite{lhaaso,km2a}. This paper introduces the design and performance of the new prototype detector for the Tibet AS surface array.

\begin{figure}[htbp]
\centering
\includegraphics[width=.8\textwidth]{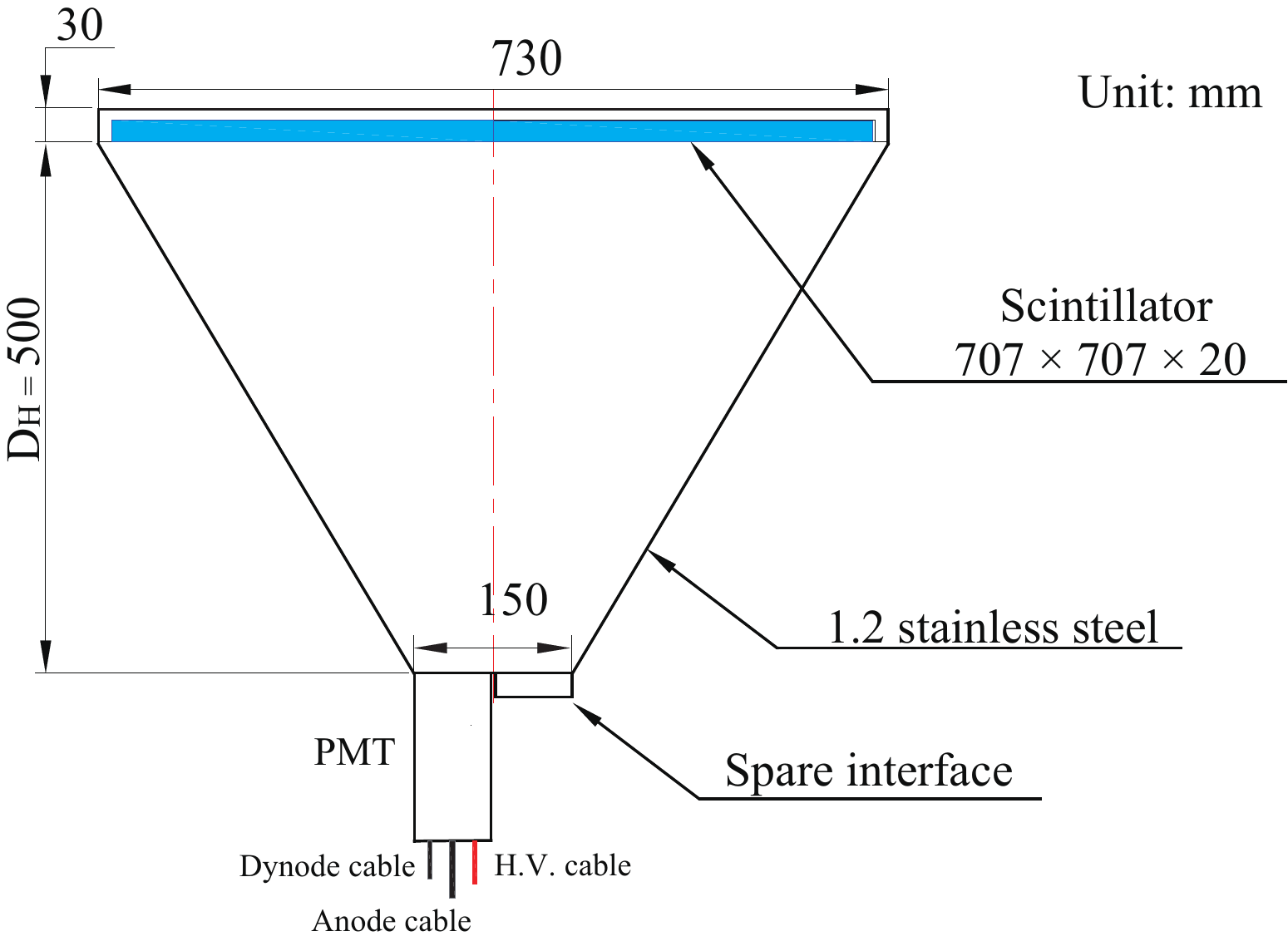}
\caption{\label{fig:a} Schematic diagram of the prototype scintillator detector.}
\end{figure}

\section{Detector design}
The scintillator detector is easy to be installed and maintained, and it can achieve a long-term stable performance as well as a good time resolution. A new prototype scintillator detector is developed for the proposed upgrading of the Tibet AS array by adopting an air light guide structure.
Figure~\ref{fig:a} shows the schematic diagram of the designed prototype. The frame structure is made of stainless steel with an upside--down pyramidal shape. The interfaces of the frame are lighttight. The reflective material, Dupont$^{\rm TM}$Tyvek$\circledR$ 1082D\footnote{http://www.dupont.com}, is glued on the inner surface of the frame to increase the efficiency of light collection significantly. An EJ200 plastic scintillator is adopted with a size of 707 mm $\times$ 707 mm $\times$ 20 mm and placed onto the top of the frame (in Figure~\ref{fig:a}). A photomultiplier tube (PMT), Hamamatsu R7725, with a diameter of 2 inches is chosen and installed onto the bottom. The detector height ($D_{H}$), from the undersurface of the scintillator to the PMT, is set to 500 mm. Next to the PMT, there is a spare interface for further extension and testing. When charged particles pass through the scintillator, photons will be generated. Some of these photons could be refracted into the air light guide, reflected inside the air light guide, and finally collected by the PMT.

\subsection{Optical surface of the scintillator}
\begin{figure}[htbp]
\centering
\includegraphics[width=.8\textwidth]{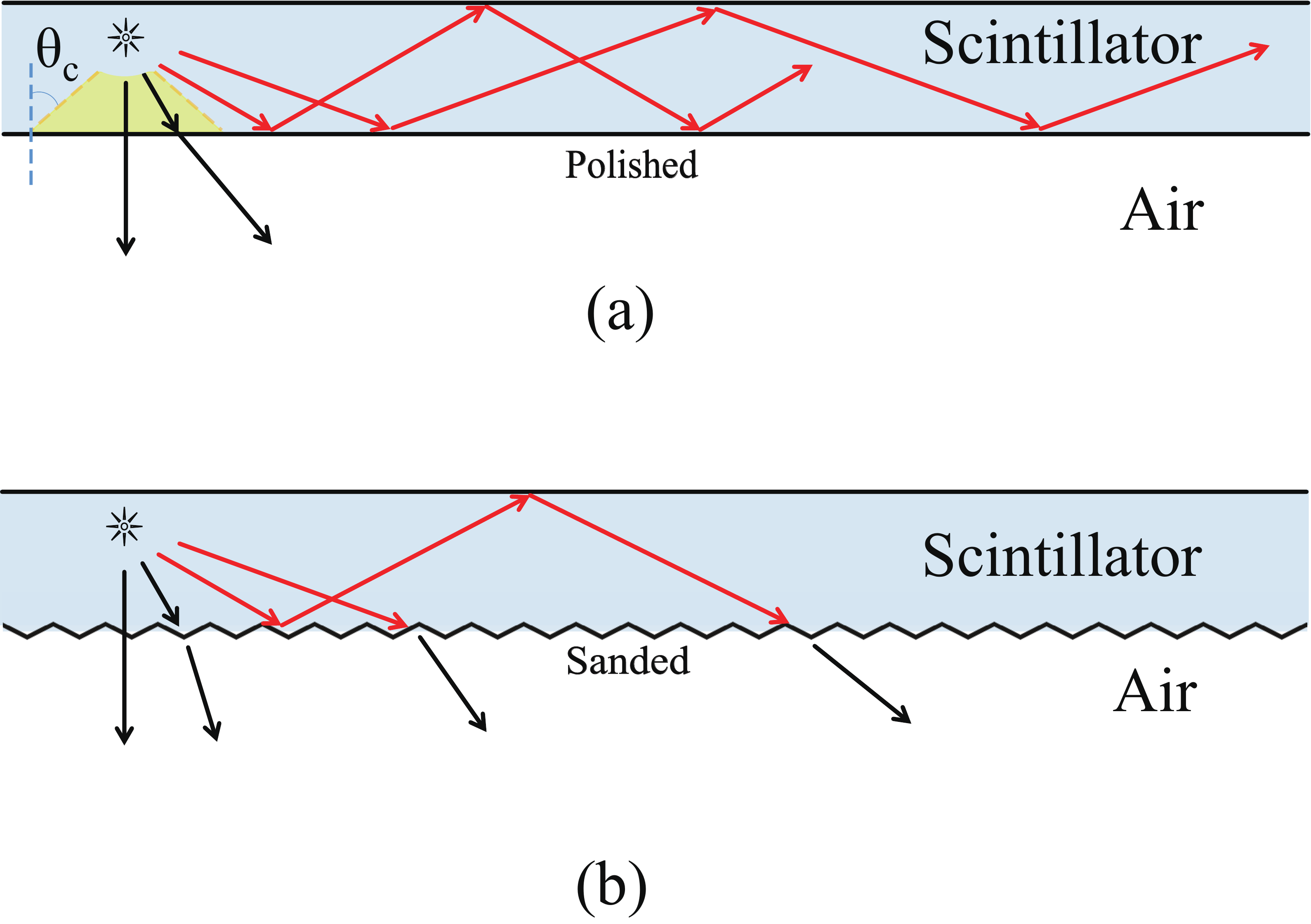}
\caption{\label{fig:b} Refraction and total internal reflection of light at the interface between the scintillator and the air light guide. (a) Two polished surfaces; (b) One surface polished and one surface sanded.}
\end{figure}

Optical photons are emitted isotropically in the scintillator. A fraction of photons will be entirely reflected when they strike the polished boundary of the scintillator at an angle larger than the critical angle (full reflection angle, $\theta_{c}$), as shown in the Figure~\ref{fig:b}-a. To reduce this total reflection effect and to increase the light output, one surface of the scintillator is sanded where the photons will be reflected and diffused in random directions, as demonstrated in figure~\ref{fig:b}-b.

In this prototype, the plastic scintillator has one polished surface, one sanded surface, and four machine-cut edges. The undersurface of the scintillator is sanded. The polished surface and four edges of the scintillator are covered with Tyvek 1082D. Compared with the two-polished-surface scintillator, the scintillator with one polished surface and one sanded surface provides an increase in light collection by $\sim$20\%.

\subsection{Design of the output circuit}
A large dynamic range and a good time resolution are two essential goals in the design of the PMT output circuit. In the Tibet AS Experiment, two PMTs are used: one fast-timing PMT is used to measure the arrival time of EAS signals, and one density PMT is added to extend the dynamic range of the detector into 500 minimum ionization particles (MIPs). In this prototype, the anode-and-dynode double readout system is adopted. A good time resolution is obtained from the anode, and a large dynamic range is achieved by adding a dynode signal. Usually, there is a good linearity between anode output currents and incident photon numbers. However, the linearity is limited by the space charge effects because of a large current flowing between the dynodes and the voltage divider circuit.

\begin{figure}[htbp]
\centering
\includegraphics[width=.95\textwidth]{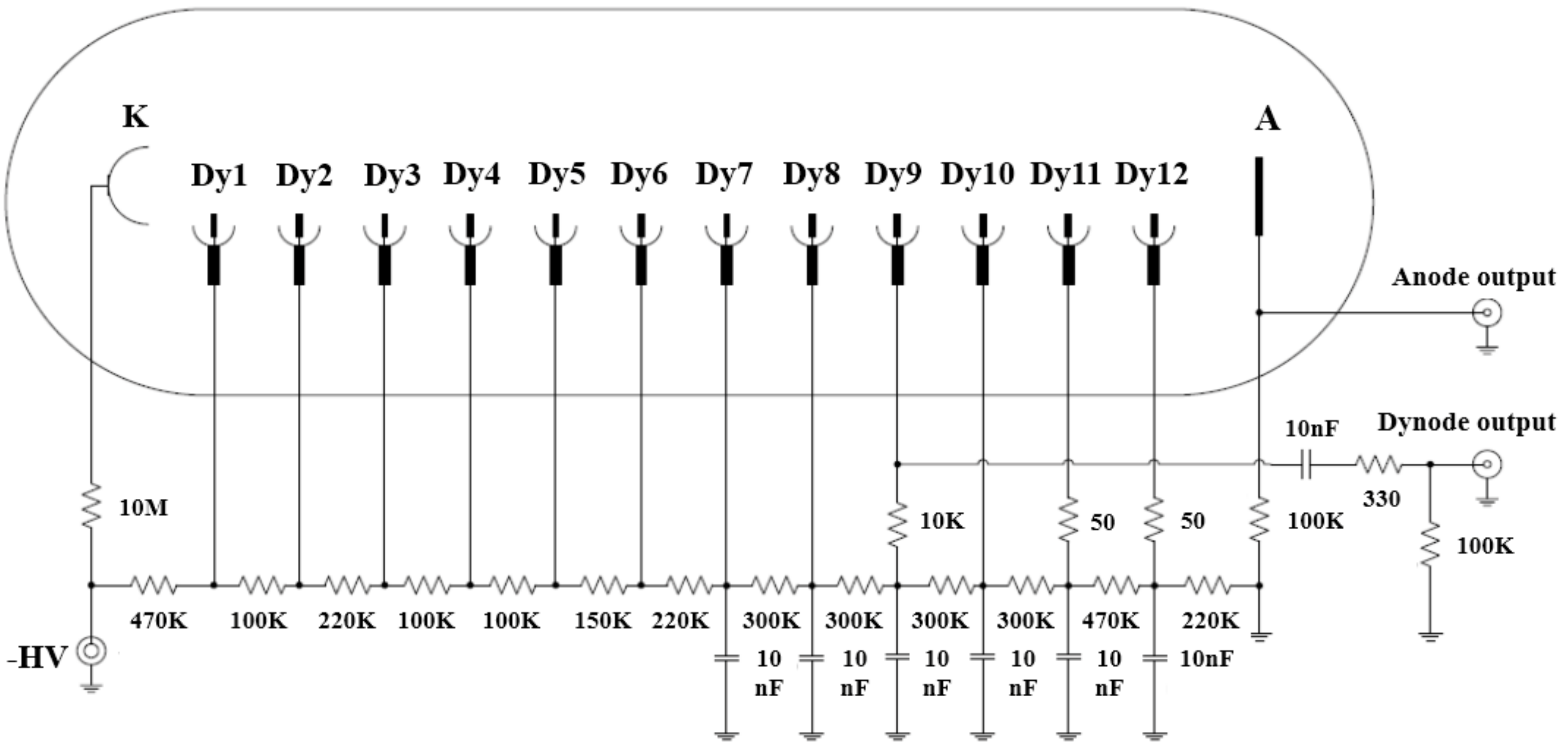}
\caption{\label{fig:c} Schematic diagram of the voltage divider circuit for PMT R7725.}
\end{figure}

In our case, PMT R7725 has 12 dynodes, and the ninth is selected as the readout to measure the large signal. The applied dynode voltages are optimized by changing resistor values of the divider, as shown in the Figure~\ref{fig:c}. With this optimized output circuit, the dynamic range from 1 to 500 MIPs has been obtained, which can fully satisfy the requirement for $\gamma$-ray observation at 100 TeV.

\subsection{Optimization of the detector geometry}
\begin{figure}[htbp]
\centering
\includegraphics[width=.75\textwidth]{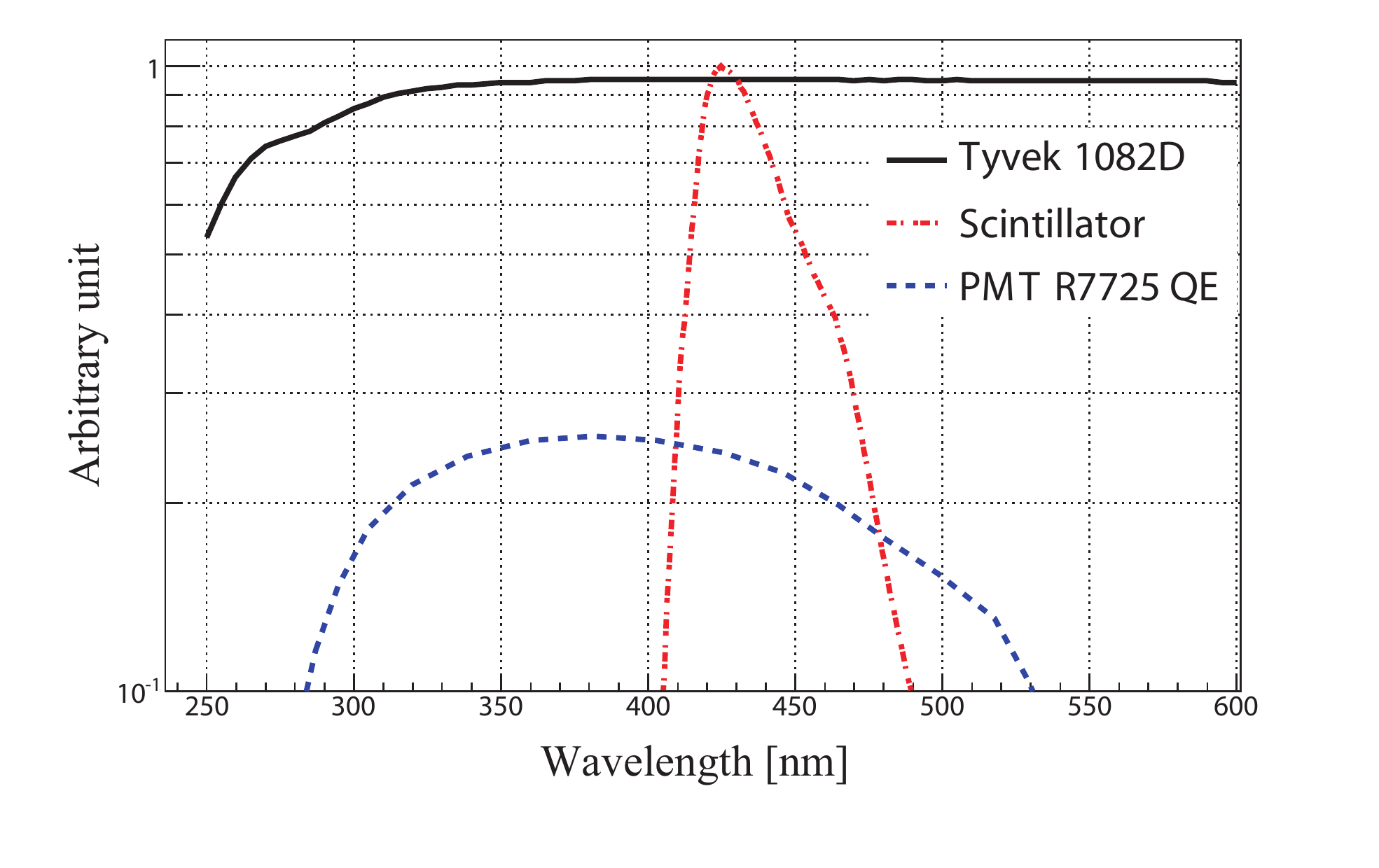}
\caption{\label{fig:d} Emission spectrum, quantum efficiency, and Tyvek 1082D reflectivity used in the Optical Model.}
\end{figure}

In order to optimize the detector geometry, a simulation package was developed based on the Geant4 toolkit \footnote{http://geant4.web.cern.ch/geant4}. In this package, the Optical Model is selected to simulate the generation, transmission and collection of optical photons. Whenever charged particles pass through the scintillator, optical photons are emitted. The generation of photons in the scintillator is characterized by its emission spectrum, as shown in the Figure~\ref{fig:d} (dashed-dotted curve), obtained from the data sheet of the EJ-200 scintillator \cite{EJ200}. The probability of the photons going out of the scintillator depends on the property of the scintillator surface, which is described by the Ground Model in the Unified Model in Geant4 \cite{Geant1996}. The propagation of an optical photon in the interface between the scintillator and the air can be described by five parameters: specular lobe reflection probability (Csl), specular spike reflection probability (Css), diffuse lobe reflection probability (Cdl), back scatter reflection probability (Cbs), and $\sigma _{\alpha}$. The values of these parameters are listed in table~\ref{tab:b}.

\begin{table}[htbp]
\centering
\caption{\label{tab:b} Parameters used in the Unified Model.}
\smallskip
\begin{tabular}{|c|c|c|c|c|c|}
\hline
   & Csl & Css & Cdl & Cbs & $\sigma _{\alpha}$ \\
\hline
Polished surface   & 98.9$\%$ & 0 & 1.1$\%$ & 0 & 0.1\\
Sanded surface     & 89$\%$ & 0 & 11$\%$ & 0 & 0.04\\
Machine-cut edges  & 96.7$\%$ & 0 & 3.3$\%$ & 0 & 0.12\\
\hline
\end{tabular}
\end{table}

\begin{table}[htbp]
\centering
\caption{\label{tab:c} Parameters of the scintillator and Tyvek 1082D used in the prototype detector.}
\smallskip
\begin{tabular}{|c|c|}
\hline
Scintllator EJ-200 & Parameters  \\
\hline
Light output/ $\%$ Anthracene           & 64\\
Scintillation efficiency/ (photons/MeV) & 10000\\
Wavelength of max. emission/ nm	        & 425\\
Rise time/ ns	                        & 0.9\\
Decay time/ ns	                        & 2.1\\
Pulse width, FWHM/ ns	                & 2.5\\
No. of H atoms/ ($10^{22}/cm^{3}$)	    & 5.17\\
No. of C atoms/ ($10^{22}/cm^{3}$)	    & 4.69\\
No. of electrons/ ($10^{22}/cm^{3}$)	& 3.33\\
Density/ (g/cc)	                        & 1.023\\
Refractive index	                    & 1.58\\
\hline
Tyvek 1082D                             & Parameters\\
\hline
Unit mass/ ($g/m^{2}$)                  & $105$\\
Thickness/ um	                        & 275\\
Opacity/ $\%$	                        & 98.4\\
Longitudinal elongation/ $\%$ (N/2.54cm)& 290\\
Transverse elongation/ $\%$ (N/2.54cm)	& 330\\
Bursting strength/ KPa	                & 1700\\
\hline
\end{tabular}
\end{table}

When most of the photons come out of the scintillator, they are reflected by high-reflective Tyvek 1082D. The Tyvek 1082D parameters are listed in table~\ref{tab:c}; its reflectivity was measured at Shandong Metrology Institute in China. After the propagation in the light-guide, only a part of these photons can be collected by the PMT, which is limited by the PMT quantum efficiency (QE) and the photoelectron collection efficiency of the first dynode. In the simulation, the emission spectrum of the scintillator, reflective efficiency of Tyvek 1082D, and quantum efficiency of PMT R7725 \cite{PMT} are shown in figure~\ref{fig:d}. Based on the parameters above, the optimization of the frame structure of the light guide is performed.

\begin{figure}[htbp]
\centering
\includegraphics[width=.8\textwidth]{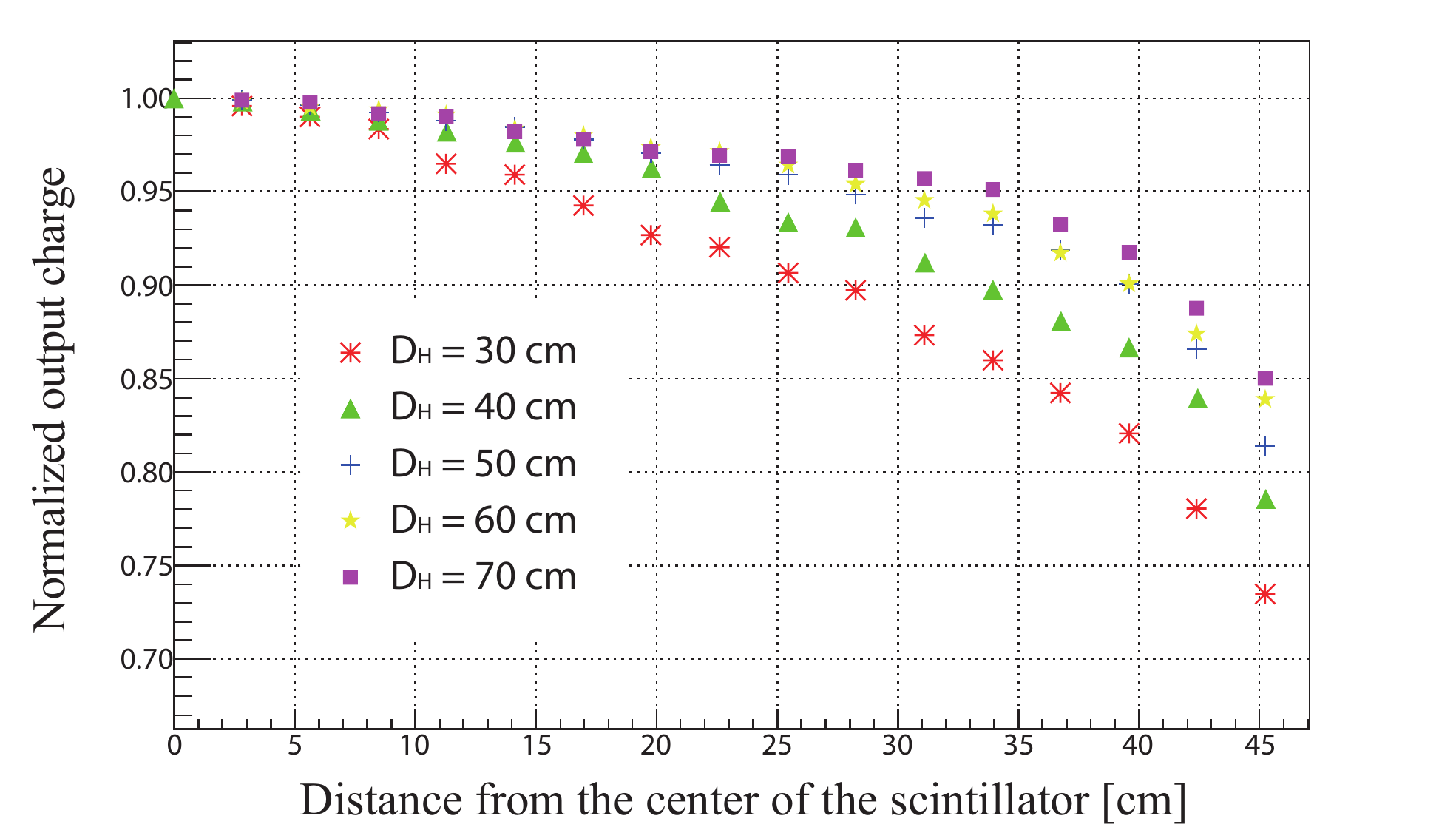}
\caption{\label{fig:e} Simulated uniformity (with reference to the center of the scintillator) at different $D_{H}$.}
\end{figure}

The frame structure of the detector is an upside--down pyramid shown in Figure~\ref{fig:a}, so the performance of the detector mainly depends on its area and height. The area of the plastic scintillator plate is fixed at 0.5 m$^{2}$ to keep consistency with the previous one in the Tibet AS$\gamma$ array. In this case, the crucial properties of the scintillator detector, such as light collection efficiency, time resolution, and positional uniformity of the output charge, are mainly determined by the height. Figure~\ref{fig:e} shows the positional uniformity at different detector heights from the MC simulation. With a height above 50 cm, non-uniformities are less than 20$\%$ with reference to the center of the scintillator. In fact, the higher the detector, the better the uniformity is, but the poorer the time resolution and the light output are. Based on this, a detector height of 50 cm is selected for the prototype.

\section{Performance of the prototype detector}
The properties of the prototype detector, such as single particle response, time resolution, uniformity, and dynamic range, are given as follows.

\subsection{Signal of the single particle}
\begin{figure}[htbp]
\centering
\includegraphics[width=.7\textwidth]{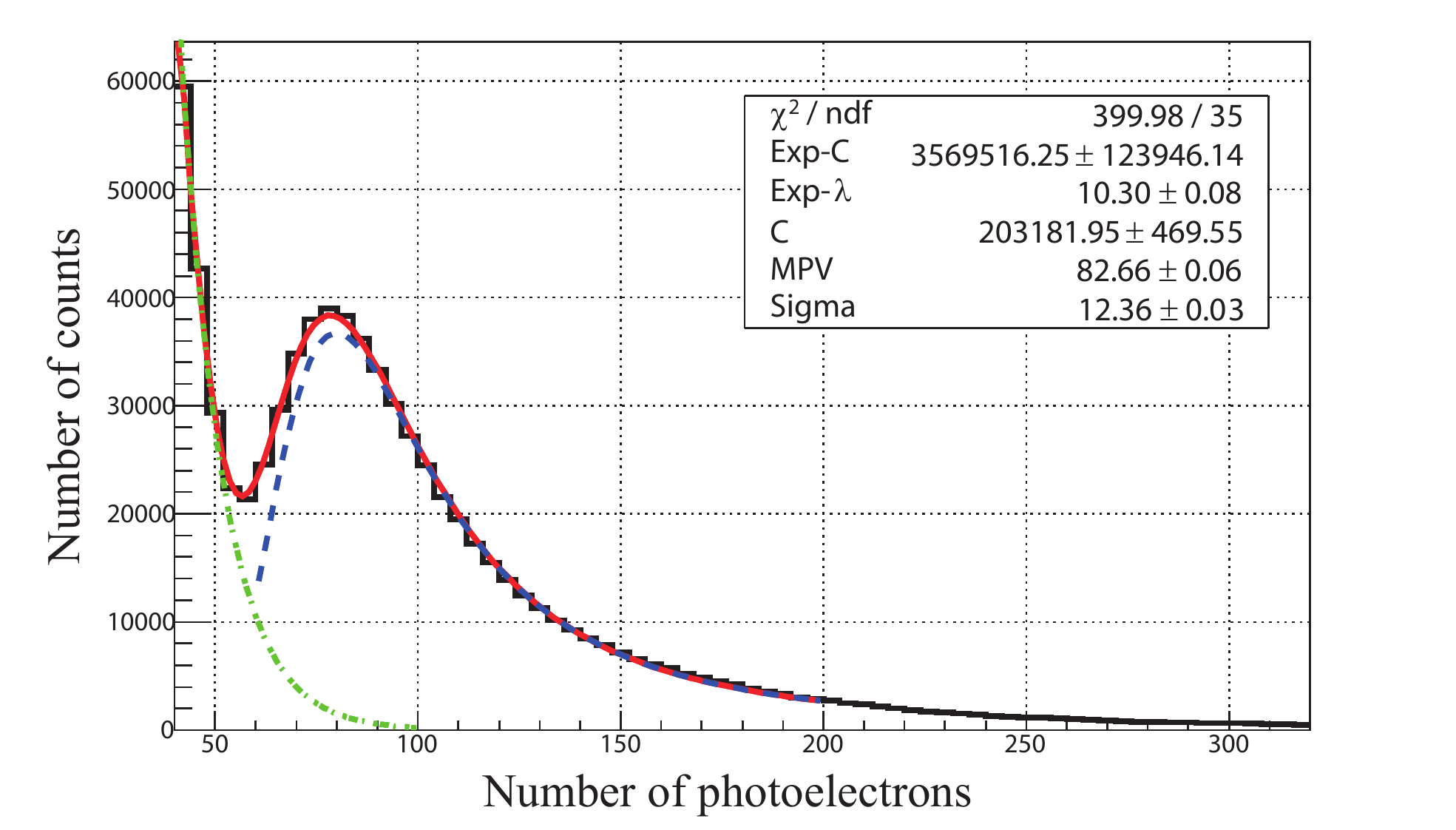}
\caption{\label{fig:f} Single-particle distribution of the prototype detector. The fitted plots include detector noise (dashed-dotted line), signal of a single particle (dashed line), and their sum (solid line).}
\end{figure}

When secondary particles produced in an air shower hit the detector and interact with the scintillator, their energy will be deposited and transformed into fluorescent light. These optical photons are then converted to electrons at the photocathode and multiplied by the dynodes of the PMT.
Cosmic muons are used to test the detector in our laboratory. The anode output signal is divided into two by a fan-in/out. One goes to the discriminator, and another one goes to the QDC. PMT R7725 works at a high voltage of 1600 V with a gain of $\sim1 \times 10^{7}$. The threshold voltage is 30 mV, corresponding to $\sim$0.25 MIPs. Figure~\ref{fig:f} shows the photoelectron distribution in the self-trigger mode. This distribution can be best fitted by a Landau function (energy deposit of single particles) plus an exponential function (contributions of the noise). The most probable value is $\sim$83 photoelectrons.

\subsection{Time resolution}
\begin{figure}[htbp]
\centering
\includegraphics[width=.7\textwidth]{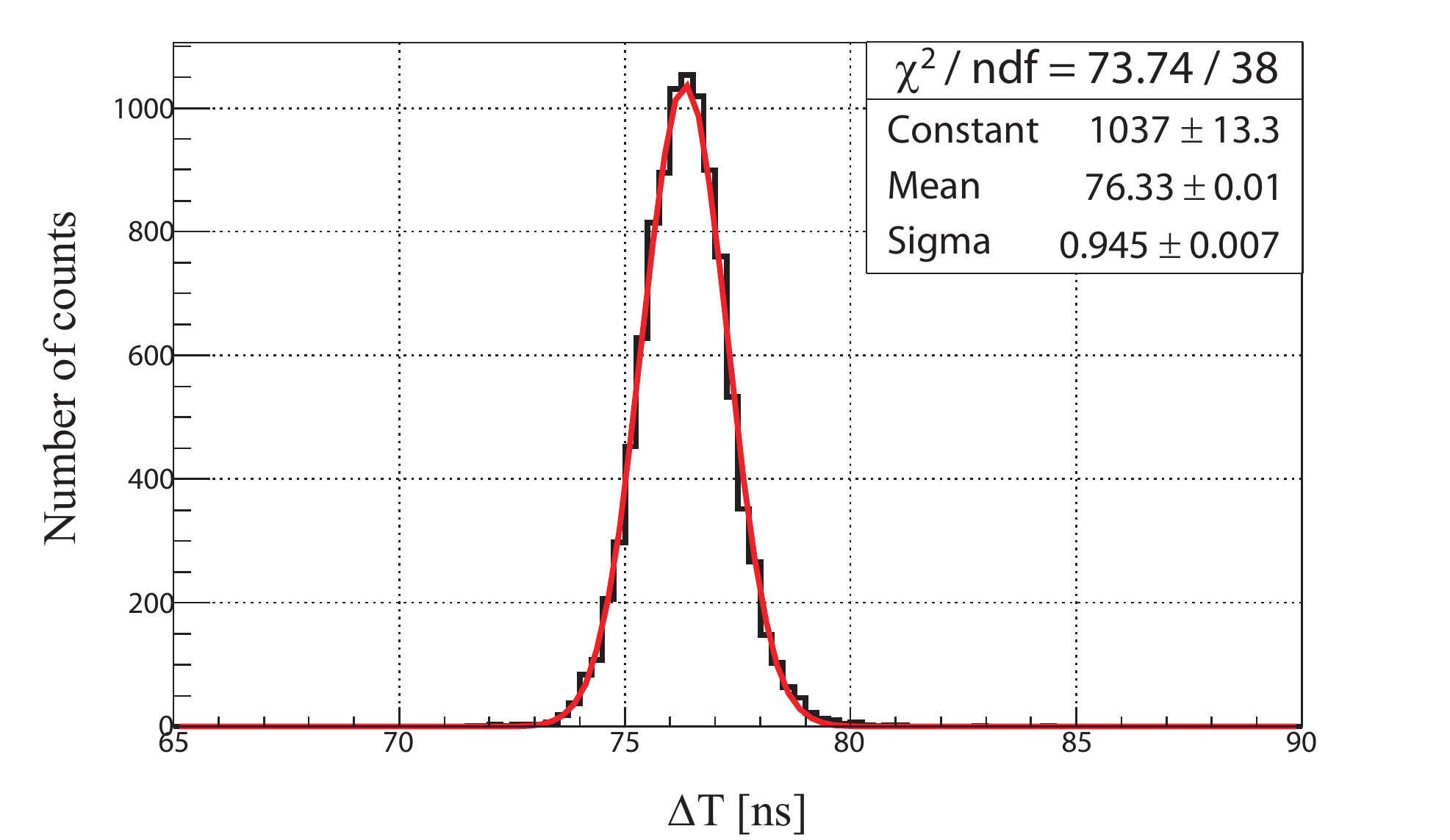}
\caption{\label{fig:g} Time resolution of the prototype detector.}
\end{figure}

For the EAS experiment, the direction of a primary CR can be inferred from the relative arrival time of secondary particles hitting the detectors. A high-precision measurement of the arrival time is very important for the direction reconstruction. To measure the time resolution of the prototype detector, a probe scintillator detector (5 cm $\times$ 5 cm $\times$ 5 cm) used as a reference was placed on the center of the prototype detector. The time resolution of the probe detector is much better than that of the prototype detector as its PMT is directly coupled to the 5-cm-thick scintillator. Coincidence events between the probe detector and the prototype detector are selected. Figure~\ref{fig:g} shows the time resolution FWHM of the prototype detector, which is $\sim$2.2 ns.

\begin{figure}[htbp]
\centering
\includegraphics[width=.65\textwidth]{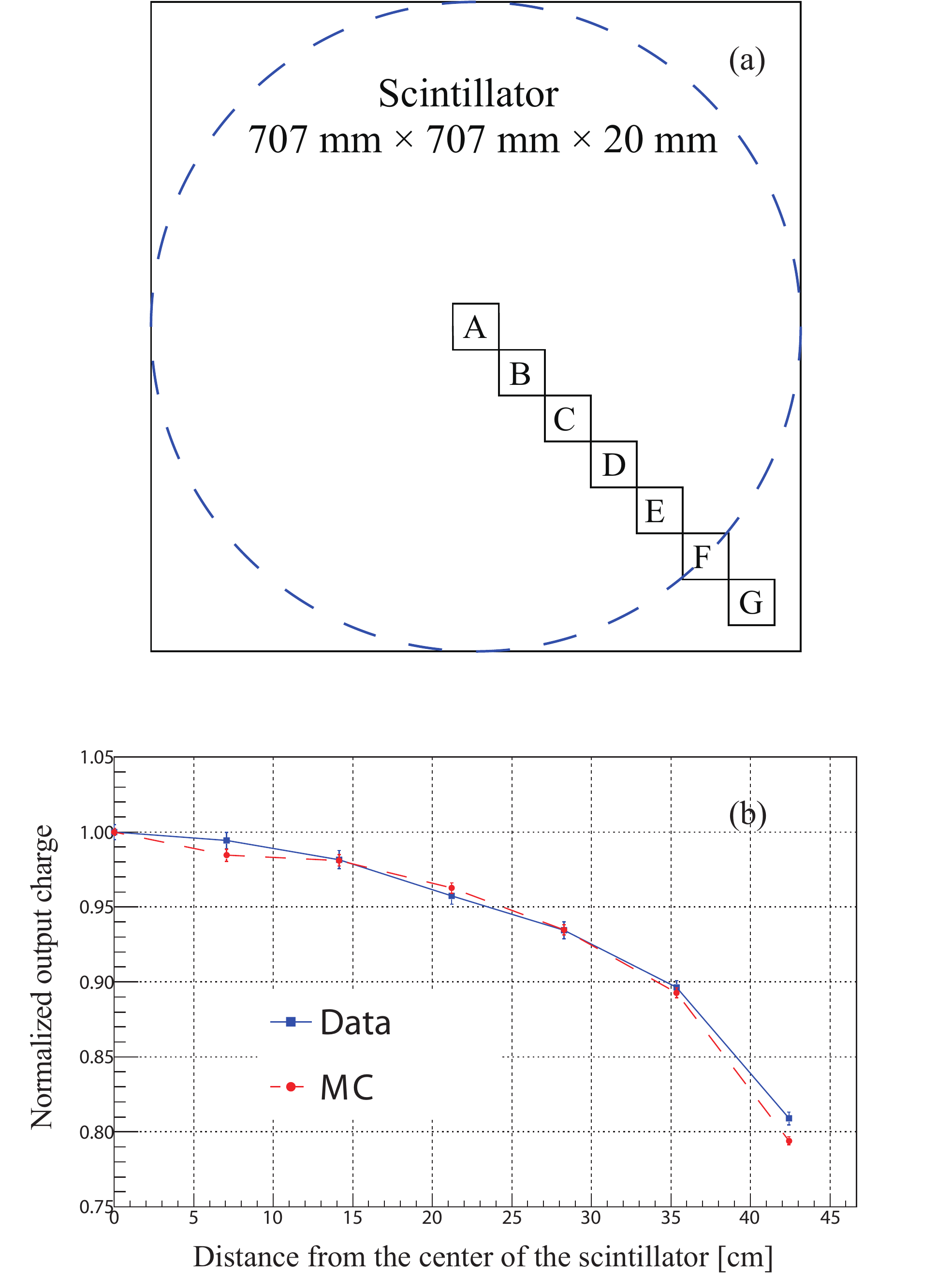}
\caption{\label{fig:h} (a) Experimental set-up for the uniformity measurement. (b) Uniformity of the prototype detector (with reference to the center of the scintillator).}
\end{figure}

\subsection{Uniformity of the output charge}
To study the positional uniformity of the output charge of the detector, the probe detector mentioned in section 3.2 was placed on 7 points along the diagonal (from center position A to edge position G, with a spacing of 5 cm) shown in figure~\ref{fig:h}-a. Figure~\ref{fig:h}-b shows the normalized output charge as a function of the position, where one can see that the data (blue squares) and the MC simulations (red triangles) described in section 2.3 are in good agreement. The non-uniformity of the output charge (from position A to F) is within 10\% with reference to the position A. On the corners (position G), the charge decreases quickly because of the edge effects.

It should be noted that only the main structures and materials are simulated in section 2.3. In the real detector, support structures are added onto four top corners to hold the plastic scintillator. These support structures can affect $\sim$3\% the efficiency of the light collection. Nevertheless, the uniformity of the prototype detector could fulfill the requirement of our experiment.

\subsection{Linearity}
The linearity of the detector mainly depends on the dynamic range of the PMT, because the plastic scintillator could not be saturated below $10^{6}$ MIPs \cite{Scin}. In this experiment, PMT R7725 has a typical linear current ($\pm$ 5$\%$ nonlinearity) of about 80 mA \cite{PMT}. To extend its dynamic range, the ninth dynode output is used as the readout to measure the large signal. With the voltage divider circuit shown in figure~\ref{fig:c}, the dynamic range of the anode output is 1-15 MIPs, while the dynode output covers the dynamic range from 5 to 500 MIPs. The overlap region can be used to calibrate the dynode gain.

\begin{figure}[htbp]
\centering
\includegraphics[width=.8\textwidth]{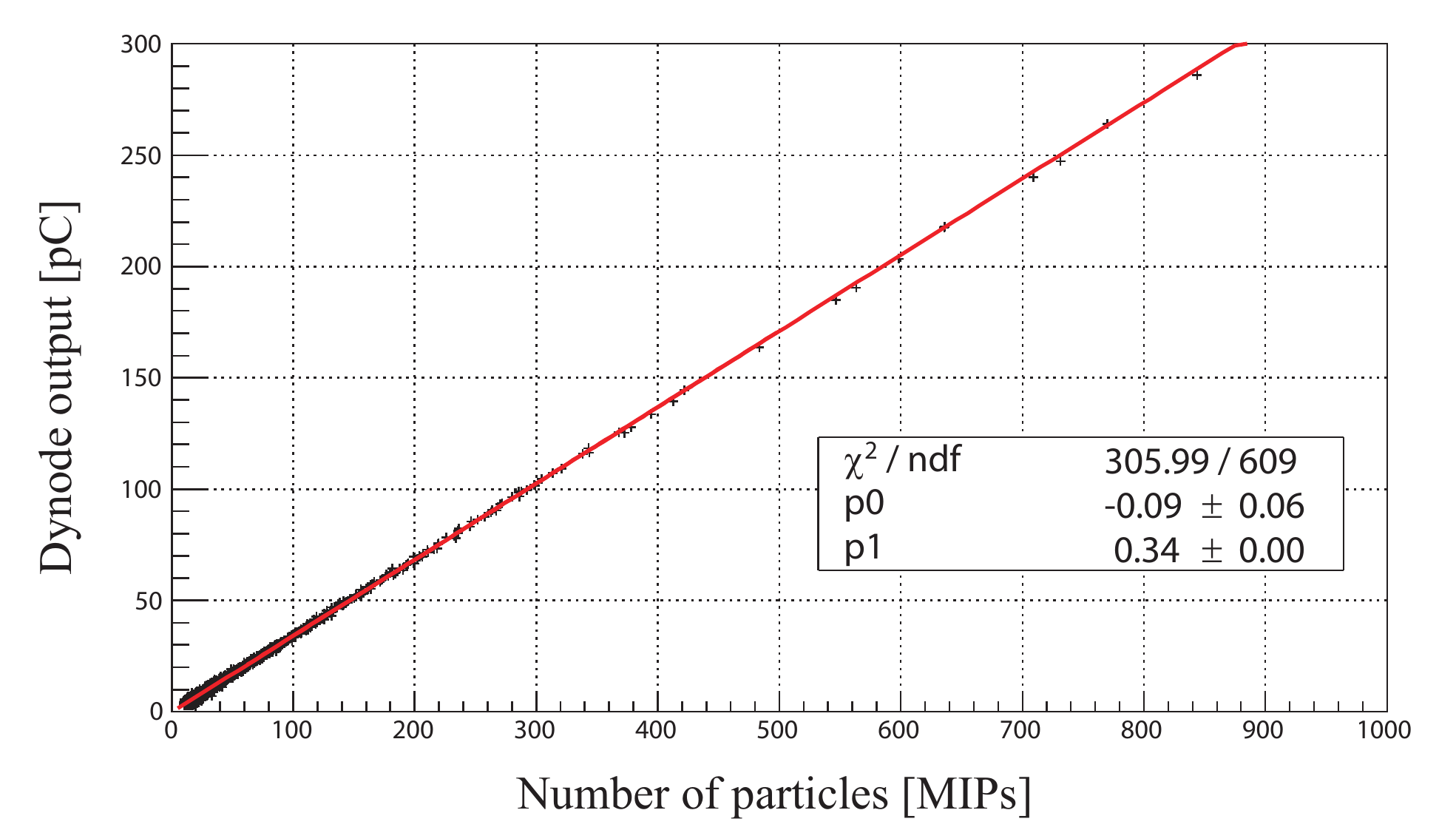}
\caption{\label{fig:i} Linearity of the prototype detector.}
\end{figure}

To test the linearity of the dynode readout, a calibrated PMT ET9829B is installed on the spare interface. The spare and the employed interfaces are symmetrically placed in the frame structure of the detector. Therefore, equal photon numbers could be collected by the two PMTs in statistically. The number of the MIPs can be estimated by the photons collected by the calibrated PMT. ET9829B's linear range could be up to 1000 MIPs, with a working voltage of 1100 V, and PMT R7725 works at a high voltage of 1240 V. Figure~\ref{fig:i} shows the dynode output of PMT R7725 as a function of the calibrated PMT's output. This linear relationship indicates that the linearity of prototype detector is good in a range from 1 to 850 MIPs.

\section{Summary}
A new air light-guide scintillator detector, with a good time resolution, moderate uniformity and dynamic range, is designed and developed to increase the effective detection area of the Tibet AS$\gamma$ array. The optimization of the detector geometry, scintillator surfaces, reflective material, and readout system was carried out. The test experiment showed that the detector's time resolution FWHM is $\sim$2.2 ns and positional non-uniformity of the output charge
within 10\% (with reference to the center of the scintillator). By using dynode readout, we can obtain the required dynamic range from 1 to 500 MIPs with one PMT.

\acknowledgments
This work is supported by Natural Science Foundation of China (11135010, 11165013, 11375209, 11405180, 11405182, 11563007, 11635011, 11663006, 11775233), 973 Program of China (2013\\CB837000), Youth Innovation Promotion Association, Chinese Academy of Sciences, Natural Science Foundation of Tibet Autonomous Region (2016ZR-TU-12), Youth Development Foundation of Tibet University (ZDPJZK1509). We thank the editor and anonymous referees for several valuable suggestions and comments, which have greatly helped to improve the paper.

\end{document}